\documentstyle[epsfig,12pt]{article}

\begin{document}

\title{{\large {\bf Supernovae, Hypernovae and Color Superconductivity}}}
\author{Deog Ki Hong\thanks{%
dkhong@pnu.edu} \\
Department of Physics\\
Pusan National University, Pusan 609-735, Korea \\
\\
Stephen D.H.~Hsu\thanks{ hsu@duende.uoregon.edu} \\ Department of
Physics \\ University of Oregon, Eugene OR 97403-5203,USA \\
\\
Francesco Sannino\thanks{%
francesco.sannino@nbi.dk}\\
NORDITA \\
Blegdamsvej 17, DK-2100 Copenhagen, Denmark}
\date{July, 2001}
\maketitle

\begin{abstract}
We argue that Color Superconductivity (CSC, Cooper pairing in quark matter
leading to the breaking of SU(3) color symmetry) may play a role in
triggering the explosive endpoint of stellar evolution in massive stars ($M
> 8 M_{\odot}$). We show that the binding energy release in the transition
of a sub-core region to the CSC phase can be of the same order of
magnitude as the gravitational binding energy release from core
collapse. The core temperature during collapse is likely below the
critical temperature for CSC, and the transition is first order,
proceeding on Fermi timescales when the pressure reaches a
critical value of several times nuclear density. We also discuss
the implications for hypernova events with total ejecta energy of
10-100 times that of type II supernova.
\end{abstract}

\newpage

\section{Introduction and Review of CSC}

Recent results have demonstrated rigorously that the ground state
of quark matter at sufficiently high density exhibits Color
Superconductivity (CSC), resulting from the Cooper pairing of
quark quasiparticles near the Fermi surface \cite{RW} --
\cite{Interface}. At asymptotic density, the nature of the order
parameter, the binding energy density and the critical temperature
are all precisely calculable. Less is known about intermediate
densities of several times nuclear density, however there are
strong indications of a Cooper pairing instability, and estimates
of the resulting gap are of order $\Delta \sim$ 40-120 MeV at a
quark chemical potential $\mu$ of $\approx$ 400 MeV. Direct links
between CSC and  the astrophysics of compact objects have been
suggested in \cite{OS}.

In this letter we discuss the possible implications of CSC for the collapse
and explosion of massive stars ($M > 8 M_{\odot}$). We argue that it is
quite likely that at the moment of maximum compression of the collapsing Fe
core, the densest part of the star crosses the critical density into the
phase where CSC is energetically favored. The release of latent energy has
the potential to generate an explosive shockwave which powers the resulting
supernova (SN). Current simulations of supernovae are generally unable to
reproduce the explosive behavior observed in nature: the shockwave generated
by the mechanical bounce of the nuclear core stalls before reaching the
surface, unless an appeal is made to additional phenomena such as neutrino
reheating combined with non-spherical phenomena such as rotation or convection
\cite{SNcollapse,SNsim}. We also note that the energy liberated in
a CSC phase transition is potentially sufficient to power hypernovae (HN)
\cite{Hyper}, which have been linked to gamma ray burst events \cite{GRB}.

First, let us summarize some results on CSC from the recent literature.
Precise results are only valid at asymptotic densities where the effective
QCD coupling is small, however they should still be useful guide when
dealing with intermediate densities. In any case, our qualitative results
will be insensitive to factors of 2 in these formulas:

\bigskip

\noindent $\bullet$ Gap size: $\Delta \sim$ 40 - 120 MeV

\noindent $\bullet$ Critical temperature: $T_c \simeq .57 \Delta$

\noindent $\bullet$ At asymptotic density the binding energy
density is $E_{CSC} = {\frac{5.8}{4 \pi^2}} \Delta^2 \mu^2$.
Recent studies \cite{Interface} of the interface between the
nuclear and CSC phase using specific models seem consistent with
this result. Simple dimensional analysis (given the absence of any
small parameter) also suggests a value in this range. Note that we
are interested here in the latent {\it energy} associated with the
first order transition to the CSC phase (or any other transition
that occurs as the baryon density is increased beyond several
times nuclear density). The baryon density changes on
astrophysical timescales, or very slowly on the timescale of QCD
dynamics, and the vacuum state at zero temperature (or at $T <<
T_c$) is found by minimizing the energy in the sector of the
Hilbert space with fixed baryon number. Studies such as
\cite{Interface}, which involve the free energy $\Omega$ at finite
chemical potential (but not fixed density) are appropriate for
determining pressure equilibrium between nuclear and CSC matter in
circumstances in which baryon number can flow across a boundary
(e.g. in a neutron star), but do not address a possible SN
transition.

\noindent $\bullet$ Phase diagram: the normal nuclear phase is
separated from the CSC phase (most likely the 2SC two flavor
condensate phase, although possibly the CFL phase \cite{Interface})
by a first order boundary.

\bigskip

\section{Astrophysics of Core Collapse}

Now let us review the standard scenario of Fe core collapse which is
believed to lead to type II supernovae \cite{SNcollapse}. Nuclear burning
during the $10^7$ year lifetime of the star leads to a shell structure, with
the inner core eventually consisting of Fe ash. Because iron cannot
participate in further exothermic nuclear reactions, there is an eventual
cooling and collapse of the Fe core, whose mass is likely to be $(1-2)~
M_{\odot}$ (or, roughly the Chandresekhar mass). The collapse of this core
is only halted by neutron degeneracy, which leads to a stiffening of the
equation of state. The resulting bounce produces a shockwave, whose energy
of $\sim 10^{51}$ ergs is a small fraction of the total available
gravitational binding energy released by the collapse:
\begin{equation}
E_b \sim 3 \cdot 10^{53} {\rm ergs} \left( {\frac{M_{core}
}{M_{\odot}}} \right)^2 \left( {\frac{R }{10 {\rm km}}}
\right)^{-1} \ . \end{equation} Most of this energy escapes in the
form of neutrinos during the supernova, as was observed in the
case of SN1987a.

The pressure in the collapsed core at the instant of the bounce is
most likely {\it greater} than the corresponding pressure in any
remnant neutron star. In order to cause a bounce, the kinetic
energy of the infalling material (which is a sizeable fraction of
a solar mass) must be momentarily stored as compressional
potential energy in the (sub)nuclear matter. This additional
mechanical squeezing at the bounce suggests that if the critical
density for CSC is ever reached in a neutron star, it will be
reached at this instant.

Simulations of the core bounce result in densities of at least
several times nuclear density ($5-10 \cdot 10^{14} {\rm g/cm^3}$),
and temperatures of roughly 10-20 MeV \cite{SNcollapse}.
This temperature is likely less than $T_c$ for CSC,
possibly much less\footnote{It is conceptually easier to think
about the case where $T$ is much less than $T_c$, since in this
case the Free energy ($F = U - TS$) liberated by the transition is predominantly
energy, with only a small component related to entropy. The relevant
dynamics is governed by energetics rather than Free energetics.}, and hence
the core of the star traverses the phase diagram horizontally in
the density-temperature plane, crossing the critical density
boundary into the CSC phase. It is important to note that the core
region at bounce is probably {\it cooler} than post-bounce,
since degenerate neutrinos tend to heat the proto-neutron star as
they diffuse out \cite{Lattimer}. Studies quoting larger SN temperatures
such as $T \sim 30$ MeV generally refer to this later stage \cite{Carter}.

Once the core crosses into the CSC part of the phase diagram, the
transition proceeds rapidly, on hadronic timescales. Because the
transition is first order, it proceeds by nucleation of bubbles of
the CSC phase in the normal nuclear background. The rate of bubble
nucleation will be of order $({\rm fm})^4$ (${\rm
fm}=10^{-13}$cm.), due to strong coupling. (In a system governed
by a dimensional scale $\Lambda$, the nucleation rate is given by
$\Gamma \sim \Lambda^4 e^{-S}$, where $S$ is the action of the
Euclidean bounce solution interpolating between the false (normal
nuclear) vacuum and a bubble of critical size. At strong coupling,
$S$ is of order one, so there is no large exponential suppression
of the nucleation rate. The scale $\Lambda$ is of the order of
$\Delta$ or $\mu = 400$ MeV.)

Causality requires that the mechanical bounce of the core happen
over timescales larger than the light crossing time of the core,
or at least $10^{-4} s$. Hence, the phase transition occurs
instantaneously on astrophysical timescales.
A nucleated bubble of CSC phase expands relativistically -- liberated latent
heat is converted into its kinetic energy -- until it collides with other
bubbles. Because the system is strongly coupled, these collisions lead to
the rapid production of all of the low energy excitations in the CSC phase,
including (pseudo)Goldstone bosons and other hadrons. The resulting release
of energy resembles an explosion of hadronic matter.

To estimate the total CSC energy released in the bounce, we use
the result that the ratio of CSC binding energy density to quark
energy density is of order $\left(\frac{ \Delta }{\mu }\right)^2$.
For $\mu \sim 400$ MeV, and $\Delta \sim$ 40-120 MeV, this ratio
is between .01 and .08, or probably a few percent.
\begin{equation}
E \sim \left(\frac{ \Delta }{\mu }\right)^2 M_{core} \ .
\end{equation}
In other words, the total energy release could be a few percent of a solar
mass, or $10^{52}$ ergs! This is significantly larger than the energy
usually attributed to the core shockwave, and possibly of the order of the
gravitational collapse energy $E_b$. The implications for SN simulations are
obviously quite intriguing.

In \cite{BH} it was suggested that strange matter formation might
overcome the energetic difficulties in producing type-II supernova
explosions. While there are strong arguments that the CSC
transition should be first order, and reasonable order of
magnitude estimates of the latent heat \cite{RW,Interface}, it is
not clear to us why there would be supercooling in a strange
matter transition. The conversion of up quarks to strange quarks
must proceed by the weak interactions, but the rate is still much
faster than any astrophysical timescale. Thus, the population of
strange quarks is likely to track its chemical equilibrium value
as the pressure of the core increases. There may be an important
effect on the nuclear equation of state from strangeness (e.g. a
softening of the pressure-density relationship), but we do not see
why there should be explosive behavior.

Our results are also relevant to hypernovae \cite{Hyper}, which
are observed to have ejecta kinetic energies 10-100 times larger
(of order $10^{52-53}$ ergs) than those of ordinary type II
supernovae. Accounting for this extra kinetic is extremely
challenging in standard scenarios. However, for exceptionally
massive stars with $M < 35  M_{\odot}$ (for $M > 35 M_{\odot}$ the
hydrogen envelope is lost during H-shell burning, and the core
size actually decreases~\cite{chlee}) there is a large core mass
which leads to a larger release of CSC binding energy. In fact,
the released energy might depend nonlinearly and sensitively on
the star's mass at the upper range. For example, the fraction of
$M_{core}$ which achieves critical density might be a sensitive
function of the mass of the star.

Another alternative is that hypernovae are the result of neutron
star mergers rather than the explosion of an individual star. This
possibility has been examined in relation to hypernovae as the engines
of gamma ray bursts (GRBs) \cite{GRB}. It seems
quite plausible that in the merger of two cold neutron stars
a significant fraction of the stars' mass undergoes the
CSC transition (i.e. crosses the critical pressure boundary for the first
time; in this case the temperature is probably negligible relative
to the CSC scale $\Delta$). This provides a substantial new source of energy
beyond gravitational binding, and may solve the ``energy crisis''
problem for this model of GRBs \cite{GRB}.

Finally, we note that the trajectory of the SN core in the
temperature-density phase diagram might be rather complicated. The
parameters suggest a density transition (at $T < T_c$), but
subsequent reheating of the core due to the explosion, or to
neutrino diffusion \cite{Lattimer} might raise the temperature
above $T_c$, and lead to additional transitions across the
temperature boundary \cite{OS,Carter}. When $T \sim T_c$, the Free
energies ($F = U - TS$) of the normal and CSC phases are
comparable, due to the larger entropy of the normal phase. The
evolution of bubbles in this regime is governed by relative Free
energies rather than energetics alone, and the transition is
presumably less explosive than the pressure transition at $T <<
T_c$.

\section{Discussion}

Our understanding of QCD at high density has evolved dramatically over the
past two years, leading to remarkable progress in understanding the QCD
phase diagram and the color superconducting state of quark matter. We have
argued here that the well-established picture of core collapse in massive
stellar evolution suggests quite strongly that CSC may play an important
role in type II supernovae, and possibly hypernovae (GRBs).

Our assumptions concerning key parameters are conservative, and taken from
distinct (and heretofore independent) regimes of inquiry: stellar
astrophysics and dense quark matter.
Yet, they point to the interesting possibility that supernova
explosions are powered by CSC binding energy. It is well established that
the shockwave energy from core collapse is insufficient to produce an
explosion, and recent results incorporating Boltzman transport of neutrinos
show that neutrino reheating is also insufficient unless non-spherical
phenomena such as rotation or convection are taken into account \cite{SNsim}.
We are optimistic that future progress in simulations will tell us much
about whether and how latent energy from color superconductivity
plays a role in stellar explosions.

\bigskip \noindent {\bf Acknowledgements}

\noindent We would like to thank I.~Bombaci, C.-H. Lee, R.~Ouyed,
and M.~Rho for helpful discussions. The work of S.H. was supported
in part under DOE contract DE-FG06-85ER40224 and by the NSF
through the Korean-USA Cooperative Science Program, 9982164. The
work of D.K.H. was supported by the KOSEF through the Korean-USA
Cooperative Science Program, 2000-111-04-2. The work of F.S. has
been partially supported by the EU Commission under contract
HPRN-CT-2000-00130.

\end{document}